\documentclass{pasj00}

\begin{document}
\SetRunningHead{T. Kato et al.}{Superhumps of CC Cancri}

\Received{}
\Accepted{}

\title{Superhumps of CC Cancri Revisited}

\author{Taichi \textsc{Kato}, Makoto \textsc{Uemura}, Ryoko \textsc{Ishioka}}
\affil{Department of Astronomy, Kyoto University,
       Sakyo-ku, Kyoto 606-8502}
\email{(tkato,uemura,ishioka)@kusastro.kyoto-u.ac.jp}

\email{\rm{and}}

\author{Jochen \textsc{Pietz}}
\affil{Rostocker Str. 62, 50374 Erftstdat, Germany}
\email{Jochen.Pietz@t-online.de}


\KeyWords{accretion, accretion disks
          --- stars: dwarf novae
          --- stars: novae, cataclysmic variables
          --- stars: individual (CC Cancri)}

\maketitle

\begin{abstract}
    We observed the 2001 November superoutburst of CC Cnc.  This observation
makes the first detailed coverage of a superoutburst of this object.
The best-determined mean superhump period is 0.075518$\pm$0.000018 d, which
is 2.7\% longer than the reported orbital period.  This fractional superhump
excess is a quite typical value for a normal SU UMa-type dwarf nova,
excluding the previously raised possibility that CC Cnc may have
an anomalously large fractional superhump excess.
During the superoutburst plateau, the object showed a decrease of
the superhump period at $\dot{P}/P$ = $-$10.2$\pm$1.3 $\times$ 10$^{-5}$,
which is one of the largest negative period derivative known in all
SU UMa-type dwarf novae.
\end{abstract}

\section{Introduction}
   Dwarf novae are a class of cataclysmic variables (CVs), which are
close binary systems consisting of a white dwarf and a red dwarf secondary
transferring matter via the Roche lobe overflow.  A class of dwarf novae,
called SU UMa-type dwarf novae, show superhumps during their long,
bright outbursts (superoutbursts).  [For a recent review of dwarf novae
and SU UMa-type dward novae, see \citet{osa96review} and
\citet{war95suuma}, respectively.]  Superhumps have periods a few percent
longer than the orbital periods (\cite{vog80suumastars};
\cite{war85suuma}), which is believed to be a consequence of the
apsidal motion of an eccentric accretion disk \citep{osa85SHexcess}.
The fractional superhump excess ($\epsilon=P_{\rm SH}/P_{\rm orb}-1$,
where $P_{\rm SH}$ and $P_{\rm orb}$ are superhump and orbital periods,
respectively) is widely believed to be an excellent measure of the mass
ratio ($q=M_2/M_1$) of the binary system both from theoretical calculations
(\cite{osa85SHexcess}; \cite{hir90SHexcess}; \cite{lub91SHa};
\cite{lub91SHb}; \cite{mur98SH}; \cite{mur00SHprecession};
\cite{woo00SH}; \cite{mon01SH})
and observations (\cite{mol92SHexcess}; \cite{min92BHXNSH};
\cite{pat98evolution}; \cite{odo00SH}).  Most of SU UMa-type systems are
on a tight relation (originally discovered by \citet{StolzSchoembs} and
extended by various authors, e.g. \cite{tho96Porb}) between $P_{\rm SH}$ and
$\epsilon$, which is considered to be a natural consequence that
most of CVs have non-evolved low-mass secondary stars (cf.
\cite{pat84CVevolution}), i.e. $M_2$ is a strong function of
$P_{\rm orb}$, which mostly determines $q$.

    Most recently, an SU UMa-type dwarf nova (1RXS J232953.9+062814:
\cite{uem01j2329iauc}) is found to conspicuously violate this relation
\citep{uem02j2329letter}.
Subsequent spectroscopy revealed that this object has a secondary star
more massive and evolved than what is expected for the orbital period
\citep{tho02j2329}.  Departures from this $P_{\rm SH}$ vs. $\epsilon$
relation are thus candidate systems with unusual stellar parameters.

    CC Cnc [see \citet{kat97cccnc} for a historical review of this object]
is one of such candidates which was reported to have a significantly
large $\epsilon$=4.9$\pm$0.5 \% \citep{tho97vzpyxcccncaheri}, who
reported $P_{\rm orb}$ = 0.07352(5) d.
Since accurate determination of the superhump period of CC Cnc
was difficult owing to unfavorable seasonal occurrences of the past
superoutbursts \citep{kat97cccnc}, a further check of the superhump
period throughout a superoutburst under favorable condition has been
absolutely needed \citep{tho97vzpyxcccncaheri}.  An excellent
opportunity arrived when the system underwent a superoutburst in 2001
November.  This outburst enabled us to for the first time follow the
entire superoutburst.  The observation started within 2.5 d of the
outburst detection by Mike Simonsen (visual magnitude 13.2 on November 10).

\section{Observation}

\begin{figure}
  \begin{center}
    \FigureFile(88mm,60mm){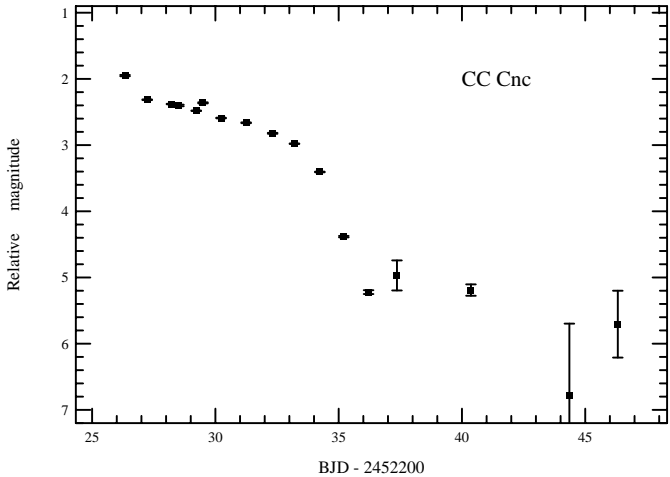}
  \end{center}
  \caption{Overall light curve of the 2001 November outburst od CC Cnc.
  The points and error bars represent averaged magnitudes and
  errors of each continuous run.}
  \label{fig:burst}
\end{figure}

    The observations were mainly done using an unfiltered ST-7E camera
attached to a 25-cm Schmidt-Cassegrain telescope at Kyoto University.
Some Kyoto observations were made using an unfiltered ST-7E camera attached
to a 30-cm Schmidt-Cassegrain telescope.  J. Pietz used an unfiltered
ST-6B camera attached to a 20-cm reflector.
All systems give magnitudes close to $R_{\rm c}$.
The exposure times were 30 s for Kyoto observations; Pietz used 60 s
and 80 s for the November 14 and 15 observations, respectively.
The images were dark-subtracted, flat-fielded, and analyzed using the
Java$^{\rm TM}$-based PSF photometry package
developed by one of the authors (TK).  The differential magnitudes of the
variable were measured against GSC 1398.1399 (averaged GSC magnitude
$V$=11.66), whose constancy during the run was confirmed by comparison
with fainter check stars in the same field.  The effect of a nearby
faint field star (cf. \cite{mis96sequence}) has been eliminated
with the PSF fitting.
The log of observations is summarized in table \ref{tab:log}.
The total number of useful frames was 5586.
Barycentric corrections were applied before the period analysis.
The overall light curve is shown in figure \ref{fig:burst}.

\begin{table*}
\caption{Log of observations.}\label{tab:log}
\begin{center}
\begin{tabular}{lccccc}
\hline\hline
Date      & BJD$^*$ (start--end) & N$^\dagger$ & Mag$^\ddagger$ &
            Error$^\S$ & Inst$^\|$ \\
\hline
2001 November 12 & 52226.309--52226.368 &  90 & 1.950 & 0.010 & 1 \\
2001 November 13 & 52227.111--52227.368 & 404 & 2.315 & 0.005 & 2 \\
2001 November 14 & 52228.075--52228.369 & 690 & 2.380 & 0.004 & 2 \\
2001 November 14 & 52228.477--52228.556 &  68 & 2.401 & 0.012 & 3 \\
2001 November 15 & 52229.090--52229.366 & 475 & 2.481 & 0.004 & 2 \\
2001 November 15 & 52229.452--52229.542 &  86 & 2.362 & 0.008 & 3 \\
2001 November 16 & 52230.075--52230.375 & 476 & 2.591 & 0.005 & 2 \\
2001 November 17 & 52231.146--52231.372 & 206 & 2.663 & 0.006 & 2 \\
2001 November 18 & 52232.273--52232.377 & 245 & 2.822 & 0.007 & 2 \\
2001 November 19 & 52233.062--52233.369 & 701 & 2.979 & 0.004 & 2 \\
2001 November 20 & 52234.108--52234.375 & 625 & 3.406 & 0.005 & 2 \\
2001 November 21 & 52235.067--52235.362 & 688 & 4.383 & 0.011 & 2 \\
2001 November 22 & 52236.058--52236.375 & 747 & 5.221 & 0.029 & 2 \\
2001 November 23 & 52237.363--52237.371 &  19 & 4.969 & 0.227 & 2 \\
2001 November 26 & 52240.359--52240.377 &  43 & 5.190 & 0.086 & 2 \\
2001 November 30 & 52244.357--52244.363 &  13 & 6.783 & 1.086 & 2 \\
2001 December  2 & 52246.323--52246.327 &  10 & 5.705 & 0.505 & 2 \\
\hline
 \multicolumn{6}{l}{$^*$ BJD$-$2400000.} \\
 \multicolumn{6}{l}{$^\dagger$ Number of frames.} \\
 \multicolumn{6}{l}{$^\ddagger$ Averaged magnitude relative to GSC 1398.1399.} \\
 \multicolumn{6}{l}{$^\S$ Standard error of the averaged magnitude.} \\
 \multicolumn{6}{l}{$^\|$ 1: Kyoto (30-cm + ST-7E),
                              2: Kyoto (25-cm + ST-7E).} \\
 \multicolumn{6}{l}{\phantom{$^\|$} 3: Pietz.} \\
\end{tabular}
\end{center}
\end{table*}

\section{Results and Discussion}

\subsection{Mean Superhump Period and Profile}\label{sec:psh}

\begin{figure}
  \begin{center}
    \FigureFile(88mm,60mm){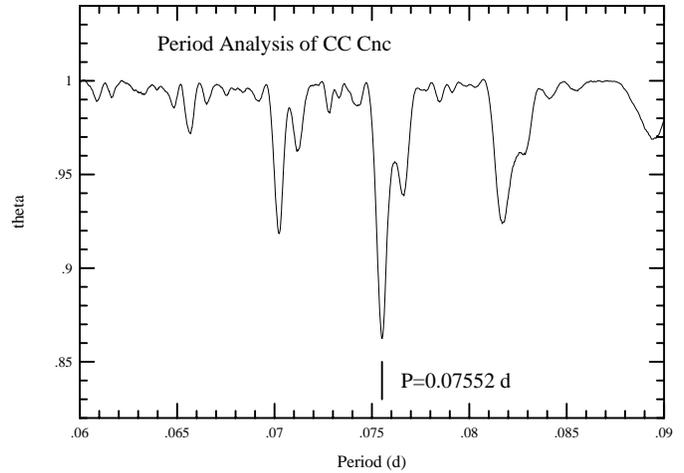}
  \end{center}
  \caption{Period analysis of superhumps in CC Cnc.  The analysis was
  done for the data between 2001 November 12 and 19 (during the superoutburst
  plateau).}
  \label{fig:pdm}
\end{figure}

\begin{figure}
  \begin{center}
    \FigureFile(88mm,60mm){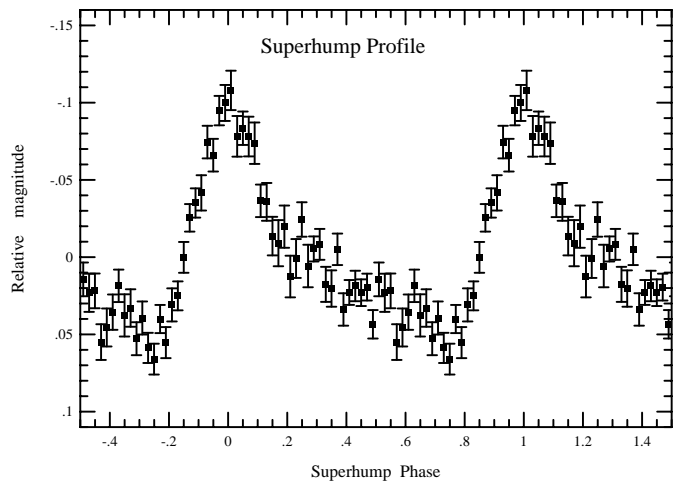}
  \end{center}
  \caption{Phase-averaged light curve of CC Cnc superhumps.}
  \label{fig:avesh}
\end{figure}

    We performed period analysis using Phase Dispersion Minimization
\citep{PDM} to all the data between 2001 November 12 and 19, after removing
the systematic trend of decline.  A correction of 0.220 mag has been
added for the 2001 November 12 data in order to correct the systematic
offset from the linear fit.  This offset was most likely a result from
a systematic difference caused by a different telescope only on this
night.
The resultant $\theta$-diagram and the phase averaged profile of superhumps
are shown in figures \ref{fig:pdm} and \ref{fig:avesh}, respectively.
The best-determined superhump period is 0.075518$\pm$0.000018 d.

\subsection{Development of Superhumps}\label{sec:shprof}

\begin{figure}
  \begin{center}
    \FigureFile(88mm,110mm){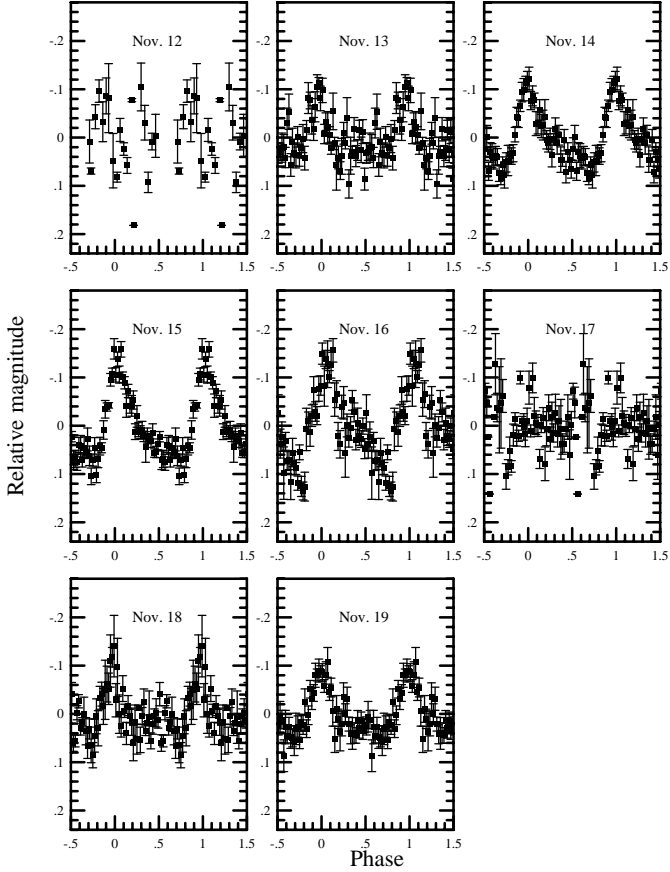}
  \end{center}
  \caption{Evolution of CC Cnc superhumps during the plateau stage of
  the superoutburst.  Each point represents an average
  of a 0.02 phase bin, except for November 12 data which used 0.04 phase bin.
  The phase zero corresponds to the zero-phase epch of equation
  \ref{equ:reg1}.  The mean superhump period (0.075518 d) was used
  to calculate the phases.
  }
  \label{fig:nightave}
\end{figure}

   Figure \ref{fig:nightave} shows nightly averaged profiles of superhumps
during the plateau stage of the superoutburst.
The amplitude of superhumps reached a maximum (0.21--0.24 mag) around
November 15--16, five days after the start of outburst.  This development
of superhumps is relatively slow compared to other SU UMa-type dwarf
novae [one of the best examples can be found in \citet{sem80v436cen};
see also \citet{vog80suumastars} and \citet{war85suuma} for general
descriptions; this delay is theoretically explained as a growth time
of the tidal instability \citep{lub91SHa}].
Although the phase coverage was not complete because of unfavorable sky
condition, the amplitude of superhumps
seems to have once decayed on November 17, and again grew on November 18.
Such a regrowth of the superhump amplitude may be related to a phenomenon
observed during the late stage of a superoutburst in V1028 Cyg
\citep{bab00v1028cyg}\footnote{
  We must note that ER UMa stars (a small, peculiar subgroup of SU UMa-type
  dwarf novae with extremely short
  supercycles; presently known members being ER UMa, V1159 Ori, RZ LMi,
  DI UMa and IX Dra (\cite{kat95eruma}; \cite{rob95eruma}; \cite{nog95rzlmi};
  \cite{kat96diuma}; \cite{ish01ixdra}) show a similar pattern of
  decay and regrowth of superhumps \citep{kat96erumaSH}.  CC Cnc, however,
  has a much longer supercycle ($\sim$400 d) than those of ER UMa stars
  (19--45 d), indicating that CC Cnc has a much lower mass-transfer rate
  than in ER UMa stars.  Although detailed mechanisms of regrowth is not
  yet identified, we consider that different mechanisms of superhump
  regrowth may be naturally taking place between ER UMa stars and other
  SU UMa-type dwarf novae.
}.
Alternately, this phenomenon seen in CC Cnc may be
also interpreted as a result of the beat phenomenon between the superhump
and orbital period (most evidently seen in eclipsing systems; e.g.
\cite{vog82zcha}; \cite{krz85oycarsuper}), as was prominently seen even
in a non-eclipsing system RZ Leo \citep{ish01rzleo}.
The calculated beat period

\begin{equation}
P_{\rm beat}={1 \over {1/P_{\rm orb}-1/P_{\rm SH}}} = 2.8~{\rm d}
\end{equation}

close to the observed time-scale of the regrowth may suggest a stronger
possibility of the second interpretation.  In this case, the orbital
inclination of CC Cnc is expected to be high, which would provide
an excellent opportunity in spectroscopically determining the component
masses and other orbital parameters.

\subsection{$O-C$ Changes}

   We determined the maximum times of superhumps from the light curve by eye.
The averaged times of a few to several points close to the maximum were
used as representatives of the maximum times.  The errors of the maximum times
are usually less than $\sim$0.004 d, which corresponds to the maximum lengths
of the data bins (i.e. a few to several points) to deduce the maximum times.
We did not use cross-correlation method to obtain individual maxima because
the profile of superhumps was rather strongly variable (subsection
\ref{sec:shprof}).  The resultant superhump maxima are
given in table \ref{tab:shmax}.  The values are given to 0.0001 d in order
to avoid the loss of significant digits in a later analysis.  The cycle count
($E$) is defined as the cycle number since Barycentric Julian Date (BJD)
2452226.322 (2001 November 12.822 UT).  A linear regression to the observed
superhump times gives the following ephemeris:

\begin{equation}
{\rm BJD (max)} = 2452226.3315 + 0.0755135 E. \label{equ:reg1}
\end{equation}

\begin{table}
\caption{Times of superhump maxima.}\label{tab:shmax}
\begin{center}
\begin{tabular}{ccc}
\hline\hline
$E$$^*$ & BJD$-$2400000 & $O-C$$^\dagger$ \\
\hline
  0 & 52226.3223 & $-$0.0092 \\
 11 & 52227.1548 & $-$0.0073 \\
 12 & 52227.2362 & $-$0.0015 \\
 23 & 52228.0679 & $-$0.0004 \\
 24 & 52228.1414 & $-$0.0024 \\
 25 & 52228.2222 &   0.0029  \\
 26 & 52228.2972 &   0.0023  \\
 29 & 52228.5276 &   0.0062  \\
 37 & 52229.1284 &   0.0029  \\
 38 & 52229.2048 &   0.0038  \\
 39 & 52229.2802 &   0.0037  \\
 42 & 52229.5049 &   0.0018  \\
 51 & 52230.1848 &   0.0021  \\
 64 & 52231.1680 &   0.0036  \\
 79 & 52232.2979 &   0.0008  \\
 80 & 52232.3738 &   0.0012  \\
 90 & 52233.1287 &   0.0010  \\
 91 & 52233.2027 & $-$0.0005 \\
 92 & 52233.2810 &   0.0023  \\
104 & 52234.1817 & $-$0.0032 \\
105 & 52234.2563 & $-$0.0041 \\
106 & 52234.3305 & $-$0.0054 \\
\hline
 \multicolumn{3}{l}{$^*$ Cycle count since BJD 2452226.322.} \\
 \multicolumn{3}{l}{$^\dagger$ $O-C$ calculated against equation
                    \ref{equ:reg1}.} \\
\end{tabular}
\end{center}
\end{table}

\begin{figure}
  \begin{center}
    \FigureFile(88mm,60mm){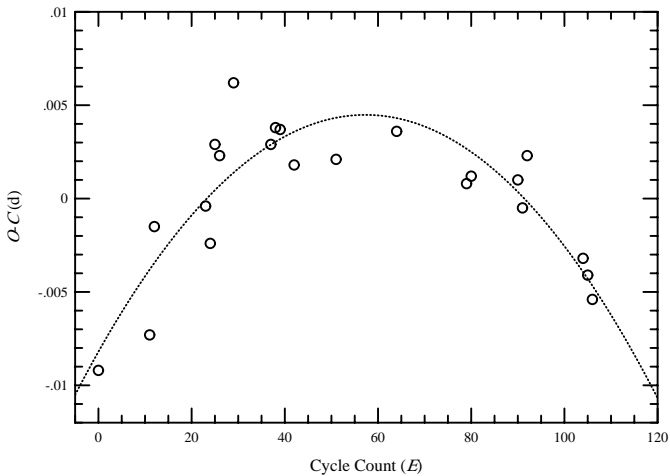}
  \end{center}
  \caption{$O-C$ diagram of superhump maxima.
  The parabolic fit corresponds to equation \ref{equ:reg2}.}
  \label{fig:oc}
\end{figure}

   Figure \ref{fig:oc} shows the ($O-C$)'s against the mean superhump period
(0.0755135 d).  The diagram clearly shows the decrease in the superhump
period throughout the superoutburst plateau.
The times of the superhump maxima in this interval
can be well represented by the following quadratic equation (the quoted
errors represent 1-$\sigma$ errors):

\begin{eqnarray}
{\rm BJD} & {\rm (max)} = 2452230.4892(7) + 0.075531(13) (E-55) \nonumber \\
    & -3.86(50) \times 10^{-6} E^2. \label{equ:reg2}
\end{eqnarray}

   The quadratic term corresponds to $\dot{P}$ = $-$7.7$\pm$1.0 $\times$
10$^{-6}$ d cycle$^{-1}$, or $\dot{P}/P$ = $-$10.2$\pm$1.3 $\times$ 10$^{-5}$.
\citet{kat01hvvir} noted that short-period systems or infrequently
outbursting SU UMa-type systems predominantly show an increase in the
superhump periods in contrast to a ``textbook" decrease of the superhump
periods in usual SU UMa-type dwarf novae.  However, observations of
period changes in long $P_{\rm orb}$ systems are relatively lacking in
the literature.  Considering that the longer $P_{\rm orb}$ systems
have larger (i.e. closer to zero) $\dot{P}/P$ \citep{kat01hvvir}, or
even virtually zero (e.g. V725 Aql: \cite{uem01v725aql}; EF Peg:
K. Matsumoto in preparation, see also \citet{kat02efpeg}), there may be
a possibility that $\dot{P}/P$ makes a minimum around the period of CC Cnc.
From a theoretical viewpoint, this decrease of superhump period is
generally attributed to decreasing apsidal motion due to a decreasing disk
radius \citep{osa85SHexcess}, or inward propagation of the eccentricity
wave \citep{lub92SH}.  It may be possible these ``intermediate period"
systems like CC Cnc enable effective propagation of the eccentricity
wave, although the possibility needs to be tested by future detailed
fluid calculations.

\subsection{Superhumps during the Rapid Decline Phase}

\begin{figure}
  \begin{center}
    \FigureFile(88mm,60mm){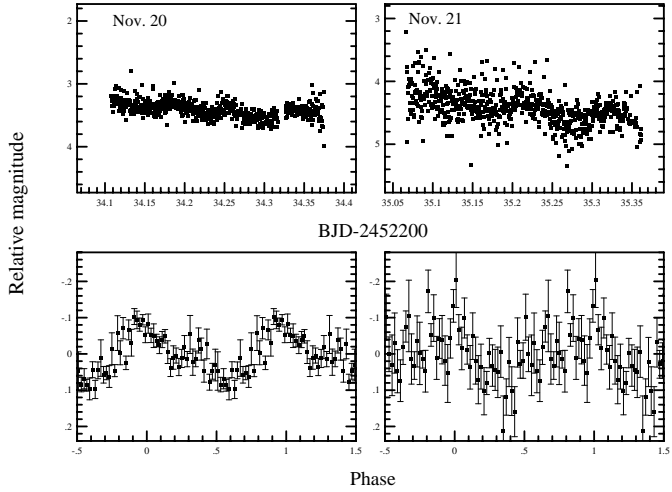}
  \end{center}
  \caption{Superhumps during the rapid decline phase (November 20--21).
  The upper panels show raw light curves.  The lower panels show averaged
  superhump profiles.  The phase-averaging follow the same prescription
  in figure \ref{fig:nightave}, after subtracting the linear decline
  trend from the raw data.
  }
  \label{fig:late}
\end{figure}

   In some SU UMa-type dwarf novae, what are called {\it late superhumps}
appear during the final stage of superoutbursts.  Late superhumps have
similar periods with ordinary superhumps (i.e. superhumps observed during
the plateau stage, subsections \ref{sec:psh}, \ref{sec:shprof}),
but have phases of $\sim$0.5 different from those of ordinary superhumps
(\cite{hae79lateSH}; \cite{vog83lateSH}; \cite{vanderwoe88lateSH};
\cite{hes92lateSH}).  Figure \ref{fig:late} shows the late-stage
evolution of superhumps in CC Cnc.  On November 20, the system started
to decline rapidly.  Ordinary superhumps were clearly present, without
a hint of $\sim$0.5 phase jump.  On November 21, the system further faded
by $\sim$1.0 mag.  Although the profile of variation became more
irregular, the maximum phase remained close to zero, suggesting that
late superhumps were weak in this system.

\section{Summary}

   We observed the 2001 November superoutburst of the SU UMa-type dwarf
nova CC Cnc.  We obtained the mean superhump period of
0.075518$\pm$0.000018 d, which is 2.7\% longer than the orbital period.
This observation excludes the previously suggested possibility that
CC Cnc may have an anomalously large fractional superhump excess.
The full growth of superhumps took $\sim$5 d from the start of the
superoutburst, which is relatively large for a long-period SU UMa-type
dwarf nova.  There was a suggestion of a regrowth of superhumps
during the late plateau stage of the superoutburst, which may be
interpreted as a result of the beat phenomenon.  During the rapid
decline stage, CC Cnc did not show prominent late superhumps.
The observed superhump period change $\dot{P}/P$ = $-$10.2$\pm$1.3
$\times$ 10$^{-5}$ is one of the largest negative period derivative
known in all SU UMa-type dwarf novae.  This may be an indication that
$\dot{P}/P$ makes a minimum around the period of CC Cnc.

\vskip 3mm

We are grateful to Mike Simonsen for promptly notifying us of the
outburst.
This work is partly supported by a grant-in aid (13640239) from the
Japanese Ministry of Education, Culture, Sports, Science and Technology.
Part of this work is supported by a Research Fellowship of the
Japan Society for the Promotion of Science for Young Scientists (MU).


\begin{thebibliography}{}

\bibitem[Baba et~al.(2000)]{bab00v1028cyg}
  Baba, H., Kato, T., Nogami, D., Hirata, R., Matsumoto, K., \& Sadakane, K.\
  2000, \pasj, 52, 429

\bibitem[Haefner et~al.(1979)]{hae79lateSH}
  Haefner, R., Schoembs, R., \& Vogt, N.\ 1979, \aap, 77, 7

\bibitem[Hessman et~al.(1992)]{hes92lateSH}
  Hessman, F.~V., Mantel, K.-H., Barwig, H., \& Schoembs, R.\ 1992, \aap, 263,
  147

\bibitem[Hirose, Osaki(1990)]{hir90SHexcess}
  Hirose, M., \& Osaki, Y.\ 1990, \pasj, 42, 135

\bibitem[Ishioka et~al.(2001a)]{ish01ixdra}
  Ishioka, R., Kato, T., Uemura, M., Iwamatsu, H., Matsumoto, K., Martin, B.,
  Billings, G.~W., \& Nov\'{a}k, R.\ 2001a, \pasj, 53, L51

\bibitem[Ishioka et~al.(2001b)]{ish01rzleo}
  Ishioka, R., Kato, T., Uemura, M., Iwamatsu, H., Matsumoto, K., Stubbings,
  R., Mennickent, R., Billings, G.~W., {et~al.}\ 2001b, \pasj, 53, 905

\bibitem[Kato(2002)]{kat02efpeg}
  Kato, T.\ 2002, \pasj, 54, 87

\bibitem[Kato, Kunjaya(1995)]{kat95eruma}
  Kato, T., \& Kunjaya, C.\ 1995, \pasj, 47, 163

\bibitem[Kato, Nogami(1997)]{kat97cccnc}
  Kato, T., \& Nogami, D.\ 1997, \pasj, 49, 341

\bibitem[Kato et~al.(1996a)]{kat96diuma}
  Kato, T., Nogami, D., \& Baba, H.\ 1996a, \pasj, 48, L93

\bibitem[Kato et~al.(1996b)]{kat96erumaSH}
  Kato, T., Nogami, D., \& Masuda, S.\ 1996b, \pasj, 48, L5

\bibitem[Kato et~al.(2001)]{kat01hvvir}
  Kato, T., Sekine, Y., \& Hirata, R.\ 2001, \pasj, 53, 1191

\bibitem[Krzeminski, Vogt(1985)]{krz85oycarsuper}
  Krzeminski, W., \& Vogt, N.\ 1985, \aap, 144, 124

\bibitem[Lubow(1991a)]{lub91SHa}
  Lubow, S.~H.\ 1991a, \apj, 381, 259

\bibitem[Lubow(1991b)]{lub91SHb}
  Lubow, S.~H.\ 1991b, \apj, 381, 268

\bibitem[Lubow(1992)]{lub92SH}
  Lubow, S.~H.\ 1992, \apj, 401, 317

\bibitem[Mineshige et~al.(1992)]{min92BHXNSH}
  Mineshige, S., Hirose, M., \& Osaki, Y.\ 1992, \pasj, 44, L15

\bibitem[Misselt(1996)]{mis96sequence}
  Misselt, K.~A\ 1996, \pasp, 108, 146

\bibitem[Molnar, Kobulnicky(1992)]{mol92SHexcess}
  Molnar, L.~A., \& Kobulnicky, H.~A.\ 1992, \apj, 392, 678

\bibitem[Montgomery(2001)]{mon01SH}
  Montgomery, M.~M.\ 2001, \mnras, 325, 761

\bibitem[Murray(1998)]{mur98SH}
  Murray, J.~R.\ 1998, \mnras, 297, 323

\bibitem[Murray(2000)]{mur00SHprecession}
  Murray, J.~R.\ 2000, \mnras, 314, 1P

\bibitem[Nogami et~al.(1995)]{nog95rzlmi}
  Nogami, D., Kato, T., Masuda, S., Hirata, R., Matsumoto, K., Tanabe, K., \&
  Yokoo, T.\ 1995, \pasj, 47, 897

\bibitem[O'Donoghue(2000)]{odo00SH}
  O'Donoghue, D.\ 2000, New Astron. Rev., 44, 45

\bibitem[Osaki(1985)]{osa85SHexcess}
  Osaki, Y.\ 1985, \aap, 144, 369

\bibitem[Osaki(1996)]{osa96review}
  Osaki, Y.\ 1996, \pasp, 108, 39

\bibitem[Patterson(1984)]{pat84CVevolution}
  Patterson, J.\ 1984, \apjs, 54, 443

\bibitem[Patterson(1998)]{pat98evolution}
  Patterson, J.\ 1998, \pasp, 110, 1132

\bibitem[Robertson et~al.(1995)]{rob95eruma}
  Robertson, J.~W., Honeycutt, R.~K., \& Turner, G.~W.\ 1995, \pasp, 107, 443

\bibitem[Semeniuk(1980)]{sem80v436cen}
  Semeniuk, I.\ 1980, \aaps, 39, 29

\bibitem[Stellingwerf(1978)]{PDM}
  Stellingwerf, R.~F.\ 1978, \apj, 224, 953

\bibitem[Stolz, Schoembs(1984)]{StolzSchoembs}
  Stolz, B., \& Schoembs, R.\ 1984, \aap, 132, 187

\bibitem[Thorstensen(1997)]{tho97vzpyxcccncaheri}
  Thorstensen, J.~R.\ 1997, \pasp, 109, 1241

\bibitem[Thorstensen et~al.(2002)]{tho02j2329}
  Thorstensen, J.~R., Fenton, W.~H., Patterson, J.~O., Kemp, J., Krajci, T., \&
  Baraffe, I.\ 2002, \apjl, 567, L49

\bibitem[Thorstensen et~al.(1996)]{tho96Porb}
  Thorstensen, J.~R., Patterson, J.~O., Shambrook, A., \& Thomas, G.\ 1996,
  \pasp, 108, 73

\bibitem[Uemura et~al.(2001a)]{uem01j2329iauc}
  Uemura, M., Ishioka, R., Kato, T., Schmeer, P., Yamaoka, H., Starkey, D.,
  Vanmunster, T., \& Pietz, J.\ 2001a, \iaucirc, 7747

\bibitem[Uemura et~al.(2002)]{uem02j2329letter}
  Uemura, M., Kato, T., Ishioka, R., Yamaoka, H., Schmeer, P., Starkey, D.~R.,
  Torii, K., Kawai, N., {et~al.}\ 2002, \pasj, 54, L15

\bibitem[Uemura et~al.(2001b)]{uem01v725aql}
  Uemura, M., Kato, T., Pavlenko, E., Baklanov, A., \& Pietz, J.\ 2001b, \pasj,
  53, 539

\bibitem[van~der Woerd et~al.(1988)]{vanderwoe88lateSH}
  van~der Woerd, H., van~der Klis, M., van Paradijs, J., Beuermann, K., \&
  Motch, C.\ 1988, \apj, 330, 911

\bibitem[Vogt(1980)]{vog80suumastars}
  Vogt, N.\ 1980, \aap, 88, 66

\bibitem[Vogt(1982)]{vog82zcha}
  Vogt, N.\ 1982, \apj, 252, 653

\bibitem[Vogt(1983)]{vog83lateSH}
  Vogt, N.\ 1983, \aap, 118, 95

\bibitem[Warner(1985)]{war85suuma}
  Warner, B.\ 1985, in Interacting Binaries, ed. P.~P. Eggelton, \& J.~E.
  Pringle (Dordrecht: D. Reidel Publishing Company), ~367

\bibitem[Warner(1995)]{war95suuma}
  Warner, B.\ 1995, \apss, 226, 187

\bibitem[Wood et~al.(2000)]{woo00SH}
  Wood, M.~A., Montgomery, M.~M., \& Simpson, J.~C.\ 2000, \apjl, 535, L39

\end{thebibliography}
\end{document}